\documentclass[prb,twocolumn,nopacs,floatfix,superscriptaddress]{revtex4-2}        

\usepackage{amssymb}
\usepackage{amsmath}       
\usepackage{graphicx}
\usepackage{hyperref}
\usepackage{color}
\usepackage{enumerate}

\begin{document}

\title{\boldmath Strong magneto-optical effects in $A$Cr$_2$O$_4$ $A$=Fe, Co spinel oxides generated by tetrahedrally coordinated transition metal ions \unboldmath}

\author{V. Kocsis}
\affiliation{Department of Physics, Budapest University of
Technology and Economics and MTA-BME Lend\"ulet Magneto-optical
Spectroscopy Research Group, 1111 Budapest, Hungary}
\affiliation{RIKEN Center for Emergent Matter Science (CEMS), Wako, Saitama 351-0198, Japan}

\author{S. Bord\'acs}
\affiliation{Department of Physics, Budapest University of
Technology and Economics and MTA-BME Lend\"ulet Magneto-optical
Spectroscopy Research Group, 1111 Budapest, Hungary}

\author{J. Deisenhofer}
\affiliation{Experimental Physics 5, Center for Electronic
Correlations and Magnetism, Institute of Physics, University of
Augsburg, 86159 Augsburg, Germany}

\author{L. F. Kiss}
\affiliation{Institute for Solid State Physics and Optics, Wigner Research Centre for Physics, Hungarian Academy of Sciences, 1525 Budapest, Hungary}

\author{K. Ohgushi}
\affiliation{Department of Physics, Graduate School of Science, Tohoku University, 6-3, Aramaki Aza-Aoba, Aoba-ku, Sendai, Miyagi 980-8578, Japan}

\author{Y. Kaneko}
\affiliation{RIKEN Center for Emergent Matter Science (CEMS), Wako, Saitama 351-0198, Japan}

\author{Y. Tokura}
\affiliation{RIKEN Center for Emergent Matter Science (CEMS), Wako, Saitama 351-0198, Japan}
\affiliation{Department of Applied Physics, University of Tokyo, Hongo, Tokyo 113-8656, Japan}

\author{I. K\'ezsm\'arki}
\affiliation{Department of Physics, Budapest University of
Technology and Economics and MTA-BME Lend\"ulet Magneto-optical
Spectroscopy Research Group, 1111 Budapest, Hungary}
\affiliation{Experimental Physics 5, Center for Electronic
Correlations and Magnetism, Institute of Physics, University of
Augsburg, 86159 Augsburg, Germany}

\begin{abstract}
Magneto-optical effects have been investigated over the
infrared--visible spectral range in $A$Cr$_2$O$_4$ ($A$ = Fe, Co) spinel
oxides with non-collinear spin orders in their ground states.
We found large magneto-optical Kerr rotation and ellipticity at the
on-site $d$-$d$ transitions of the $A^{2+}$ ions located within the
charge gap. The magneto-optical Kerr rotation of
$\vartheta_{\rm Kerr}\approx$\,12\,deg observed in CoCr$_2$O$_4$ is
unprecedentedly large among magnetic semiconductors and points towards
the uniqueness of tetrahedrally coordinated Co$^{2+}$ ions in
generating strong magneto-optical response. Criteria of strong
magneto-optical effects emerging at on-site $d$-$d$ transitions of
transition metal ions are discussed.
\end{abstract}

\maketitle

\section{Introduction}
Optical rotators and isolators, produced from materials with strong
Faraday effect, are important building blocks of optical fiber
networks. The Faraday effect, that is the polarization rotation of
light traveling through a magnetic media, is induced by the
magneto-circular birefringence. The magneto-optical Kerr effect
(MOKE), another manifestation of the magneto-circular birefringence
and dichroism, is the polarization change upon the reflection of
light from the surface of magnetic
materials.\cite{Belotelov2011,Ebert1996,Sugano2000,Shimizu2001}
The MOKE provides a figure of merit for the magnetic control of
light polarization due to the simplicity and small requirements towards the 
optical elements utilizing this phenomenon.

Principal 
 material parameters governing the magnitude of the
magneto-optical effects are the spin polarization of the
photo-excited states and the strength of the spin-orbit coupling (SOI)
acting on these states. In itinerant
magnets, the MOKE can be enhanced in the vicinity of the plasma
resonance,\cite{Feil1987,Bordacs2010} reaching
magneto-optical Kerr rotation as large as $\vartheta_{\rm Kerr}\approx$
90\,deg found in CeSb.\cite{Pittini1996,Reim1986} In such cases the
large MOKE effects usually arise from moderate values of
the off-diagonal conductivity, which are highly
enhanced by the strong dispersion of the reflectivity in the
plasma-edge region and the magnetic field induced shift of the
plasma edge.\cite{Feil1987,Bordacs2010} In insulating magnets,
strong magneto-optical effects have been observed in resonance with
on-site $d$-$d$ or $f$-$f$ optical excitations of magnetic
ions.\cite{Gridnev1997,Ohgushi2008,Popova2007,Popova2007b,Bleaney1953,Bowers1955}
Contrary to itinerant magnets, here the enhanced MOKE is mainly attributed to
the strong spin-orbit coupling as the optical 
resonances are located in the optical gap.
Recently, chromium spinel oxides and chalcogenides exhibiting large
magnetocapacitance,\cite{Hemberger2005,Weber2006}
magneto-optical\cite{Ohgushi2005,Ohgushi2008} and
magneto-elastic\cite{Bordacs2010,Kocsis2013} effects have attracted
much interest.

At room temperature, FeCr$_2$O$_4$ and CoCr$_2$O$_4$ have the normal
cubic spinel structure belonging to the $Fd\overline{3}m$ space
group.\cite{Tanaka1966,Menyuk1964,Lawes2006} The structural unit cell of $A$Cr$_2$O$_4$
spinels contains two $A^{2+}$ ions, which constitute a
diamond sublattice with tetrahedral oxygen coordination,
while the Cr$^{3+}$ ions form a pyrochlore lattice.
Due to the orbital degeneracy of Fe$^{2+}$ ions in the cubic spinel
structure, FeCr$_2$O$_4$ undergoes a cooperative Jahn-Teller
distortion at
$T_{\rm JT}$=135\,K,\cite{Tanaka1966,Tsuda2010,Bordacs2010,Kocsis2013}
which reduces the crystal symmetry to tetragonal with
the $I4_1/amd$ space group.\cite{Tanaka1966} In contrast, in
CoCr$_2$O$_4$ with no orbital degree of freedom, no measurable
distortion of the cubic structure has been detected down to
$T$=10\,K.\cite{Bordacs2010,Kocsis2013}

At low temperatures both spinel oxides show non-collinear ferrimagnetic order.\cite{Ohgushi2008} According to neutron scattering
measurements on CoCr$_2$O$_4$, the ferrimagnetic structure
arising below $T_{\rm C}\approx$\,93\,K is followed by an incommensurate transverse conical spin
re-ordering on both the Co and Cr sublattices at $T_{\rm S}$=26\,K,
which becomes commensurate at $T_{\rm
lock-in}$=13\,K.\cite{Tomiyasu2004} Below its ferrimagnetic ordering
at $T_{\rm C}\approx$70\,K, FeCr$_2$O$_4$ was 
also reported to develop a conical spin order at
$T_{\rm S}$=35\,K,\cite{Shirane1964} although the details of this
magnetic structure has not been investigated yet. In FeCr$_2$O$_4$, the
ferrimagnetic transition is accompanied by further reduction of
the symmetry to orthorhombic due to magneto-elastic
effects.\cite{Ishibashi2007,Bordacs2010,Kocsis2013}

In this work, we study the MOKE in CoCr$_2$O$_4$ and FeCr$_2$O$_4$ and report
large magneto-optical Kerr rotation emerging at the on-site $d$-$d$
transitions of the tetrahedrally coordinated magnetic $A^{2+}$ ions. To
the best of our knowledge, the magneto-optical Kerr rotation of
$\vartheta_{\rm Kerr}\approx$ 12\,deg observed in CoCr$_2$O$_4$ for
photon energies close to 0.78\,eV (close to be 1.55\,$\mu$m widely used
in telecommunication) is the highest ever observed in magnetic
semiconductors and insulators.

\section{Experimental details}

Single crystals of CoCr$_2$O$_4$ used in the
present study were grown by the chemical vapour transport method as presented in Ref.~\onlinecite{Ohgushi2008}.
FeCr$_2$O$_4$ single crystals were obtained by the floating zone technique.
FeO and Cr$_2$O$_3$ were mixed and sintered in argon atmosphere for 12\,h at 1200$^{\circ}$C temperature, then after regrinding, the resultant was pressed in rod shape and further heated in Ar+1$\%$H$_2$ atmosphere for 12\,h at 1300$^{\circ}$C.
Single crystals of FeCr$_2$O$_4$ were formed in Ar+0.1$\%$H$_2$ atmosphere at pressure of 2\,atm in a halogen-incandescent lamp floating zone furnace (SC-N35HD, NEC).

    \begin{center}
    \begin{figure*}[t]

    \includegraphics[width=18.truecm]{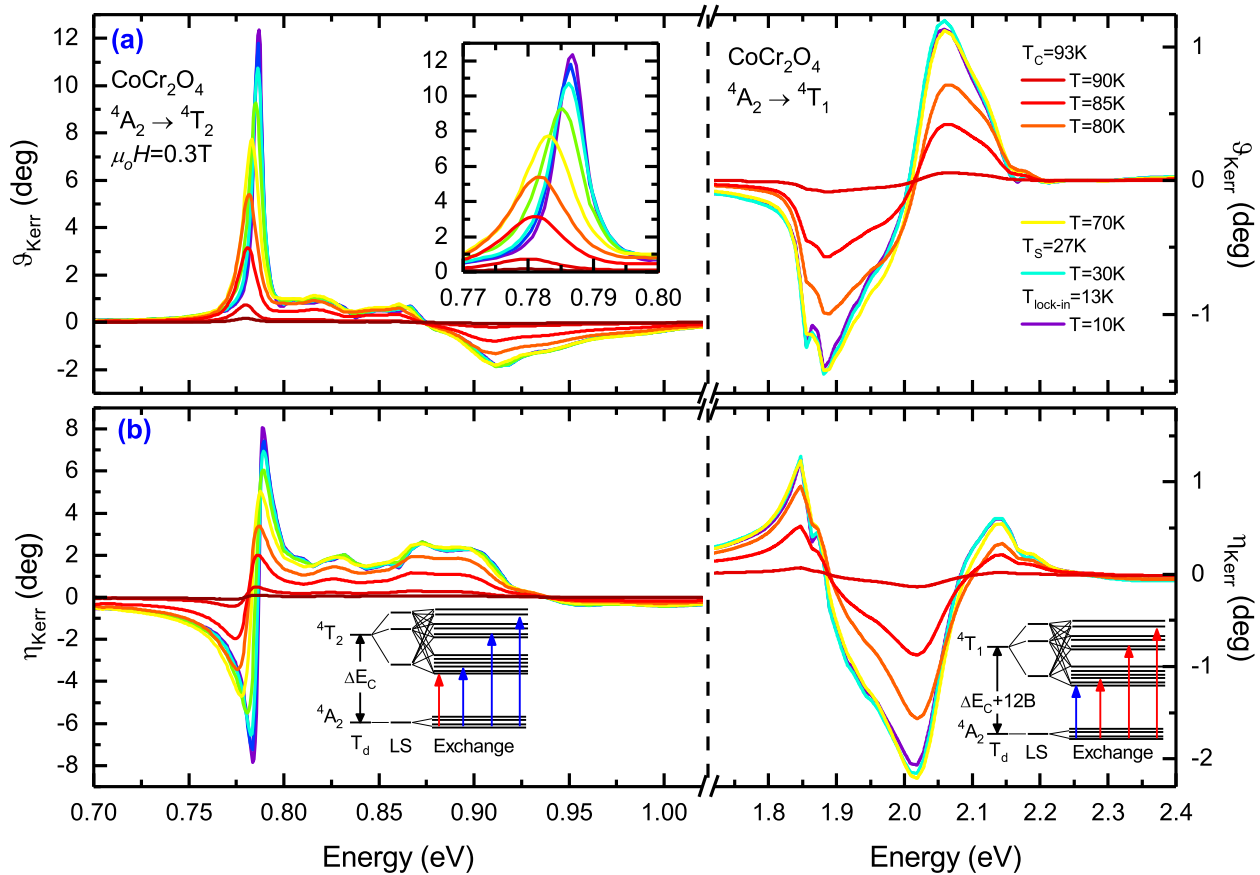}
    \caption{(Color online) Temperature dependence of the complex magneto-optical Kerr rotation spectra of CoCr$_2$O$_4$ over the near infrared--visible region measured in $\mu_oH=$0.3\,T magnetic field.
    (a)/(b) Magneto-optical Kerr rotation ($\vartheta_{\rm Kerr}$)/ellipticity ($\eta_{\rm Kerr}$) spectra at the $^4$A$_2\rightarrow^4$T$_2$ and $^4$A$_2\rightarrow^4$T$_1$ transitions of the Co$^{2+}$ ions.
    Note the breaks of the horizontal axes between the two transitions and the different vertical scales used at the two sides of the breaks. A huge peak is observed in Kerr
    rotation spectrum at the low-energy side of the $^4$A$_2\rightarrow^4$T$_2$ transition with a maximum of about
    12\,deg at $T$=10\,K. See the inset in panel (a) for an enlarged view of the peak in $\vartheta_{\rm Kerr}$. This peak is accompanied with a strong dispersive signal in the Kerr ellipticity.
    The splitting of the Co$^{2+}$ $3d$ multiplets is schematically reproduced from Refs.~\onlinecite{Sugano1970Book} and
\onlinecite{Ohgushi2008} with the following hierarchy of the
interactions: i) crystal-field of the O$_4$ tetrahedron surrounding
the Co ion, ii) spin-orbit interaction and iii) exchange
interaction. Blue (red) arrows correspond to the
$J_z \rightarrow J_z$+1 ($J_z \rightarrow J_z$-1) transitions
excited by right (left) circularly polarized photons. $J_z$ denotes
the angular momentum of the $3d$ electrons along the quantization
axis, i.e. the direction of the external magnetic field.}
    \label{moke02}
    \end{figure*}
    \end{center}

Temperature dependent reflectivity spectra were measured and
diagonal component of the optical conductivity was
calculated via Kramers-Kronig transformation as reported in a previous study.\cite{Kocsis2013} The complex Kerr rotation, $\Phi_{\rm Kerr}$, was
measured at nearly normal incidence on the (001) surface of the
crystals by a broadband dual-channel magneto-optical spectrometer as described
in earlier papers.\cite{Bordacs2010,Demko2012,Sato1981} Magnetic fields of
$\mu_oH$=$\pm$0.3\,T were applied along the propagation vector of the
incident light by permanent magnets ensuring the polar MOKE
geometry. The off-diagonal component of the optical conductivity
$\sigma_{xy}$ was evaluated using the diagonal optical conductivity
$\sigma_{xx}$ and the complex Kerr rotation, i.e. the Kerr rotation
$\vartheta_{\rm Kerr}$ and the Kerr ellipticity
$\eta_{\rm Kerr}$, according to the relation:\cite{Sugano2000,Antonov2004Book}
    \begin{equation}
    \Phi_{\rm Kerr} = \vartheta_{\rm Kerr}+i \eta_{\rm Kerr}=\frac{\sigma_{xy}}{\sigma_{xx} \sqrt{1+\frac{i\sigma_{xx}}{\omega}}}. \label{eq_01}
    \end{equation}
Temperature dependent magnetization was measured with a SQUID magnetometer 
(MPMS-5S, Quantum Design) in $\mu_oH$=0.3\,T, in accord with the MOKE experiments.

    \begin{center}
    \begin{figure}[h!]

    \includegraphics[width=8truecm]{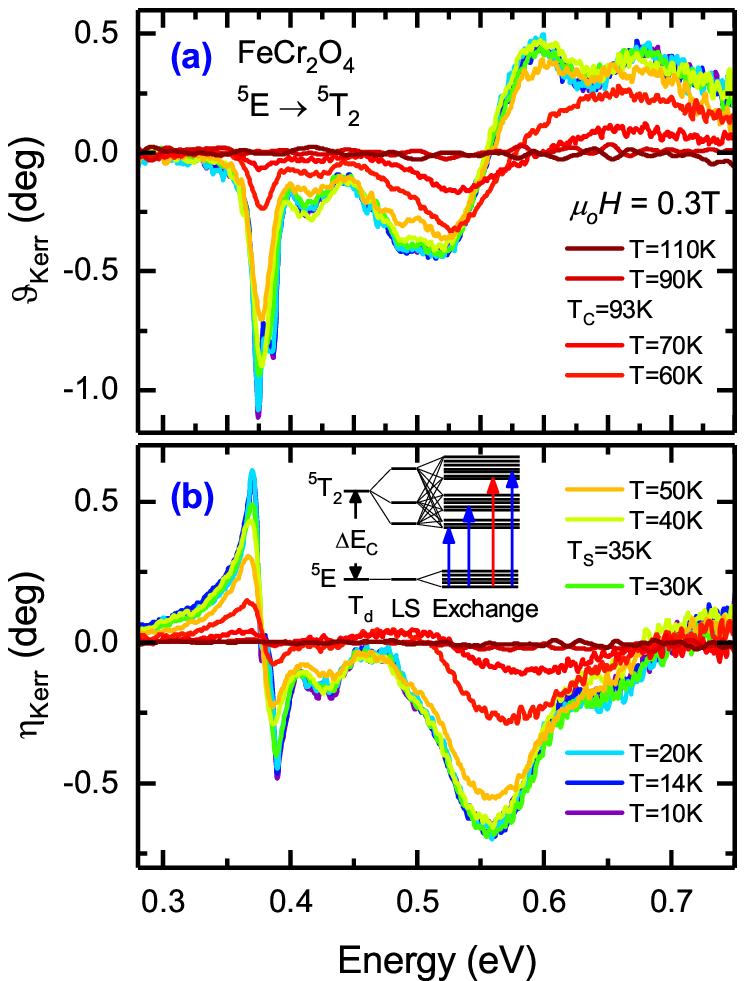}
    \caption{(Color online)Temperature dependence of the complex magneto-optical Kerr rotation spectra of FeCr$_2$O$_4$ measured in the mid and near infrared photon energy region.
    Correspondingly to Fig.~\ref{moke02}, the magneto-optical Kerr rotation (a) and ellipticity (b) were measured in $\mu_oH=\pm$0.3\,T magnetic field. Similarly to CoCr$_2$O$_4$,
    the largest Kerr rotation signal appears at the low-energy side of the $^5$E$\rightarrow^5$T$_2$ transition of the Fe$^{2+}$ ion.
    Splitting of the Fe $3d$ orbitals is schematically reproduced from Refs.~\onlinecite{Sugano1970Book} and \onlinecite{Ohgushi2008} in the same way as in Fig.~\ref{moke02}.}
    \label{moke01}
    \end{figure}
    \end{center}

\section{Results and discussion}

\subsection{Magneto-optical Kerr rotation spectra}

The temperature dependence of the complex Kerr rotation spectra of
CoCr$_2$O$_4$ and FeCr$_2$O$_4$ in the near infrared--visible region
are respectively shown in Figs.~\ref{moke02} and \ref{moke01}. In
the ferrimagnetic state of the materials, strong MOKE is found in the
energy range of the $d$-$d$ excitations of the magnetic $A^{2+}$ ions.
The assignment and the fine structure of the transitions are reproduced from Refs.~\onlinecite{Sugano1970Book} and
\onlinecite{Ohgushi2008} in the inset of the figures. In the ferrimagnetic 
phase, strong and sharp peaks emerge in the Kerr rotation spectra at the
low-energy sides of each $d$-$d$ transitions of both materials.
The low-energy side peak of the Co$^{2+}$ ions located at about
0.78\,eV, reaches the value of $\vartheta_{\rm Kerr}=$12\,deg
at $T$=10\,K. This excitation exhibits a similarly strong ellipticity
signal with a dispersive line shape. For further analysis of the
spectral features off-diagonal optical conductivity spectra were calculated
 according to Eq.~(\ref{eq_01}).

    \begin{center}
    \begin{figure*}[t]

    \includegraphics[width=18truecm]{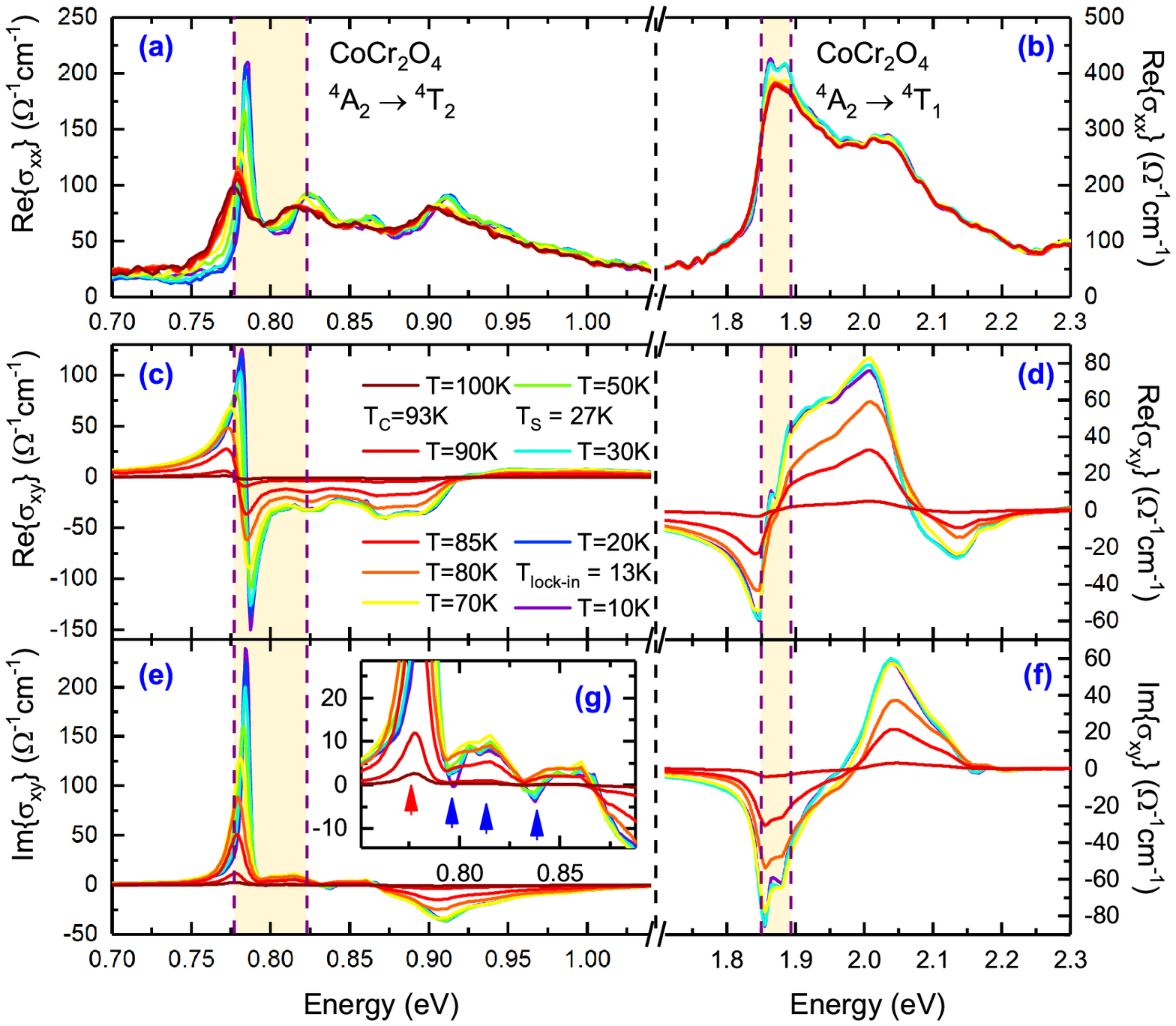}
    \caption{(Color online) Diagonal (a,b) and the complex off-diagonal (b-f) optical conductivity spectra at various temperatures of the two transitions of CoCr$_2$O$_4$. Regions of the electronic states approximated by the free-ion spin-orbit splitting for the $^4$T$_2$ and $^4$T$_1$ multiplets are indicated by the coloured stripes around 0.8\,eV and 1.9\,eV, respectively. Blue and red arrows indicate the possible arrangement of the SOI and Zeemann split ZPLs for the $^4$T$_2$ multiplet.}
    \label{moke04}
    \end{figure*}
    \end{center}

    \begin{center}
    \begin{figure}

    \includegraphics[width=8truecm]{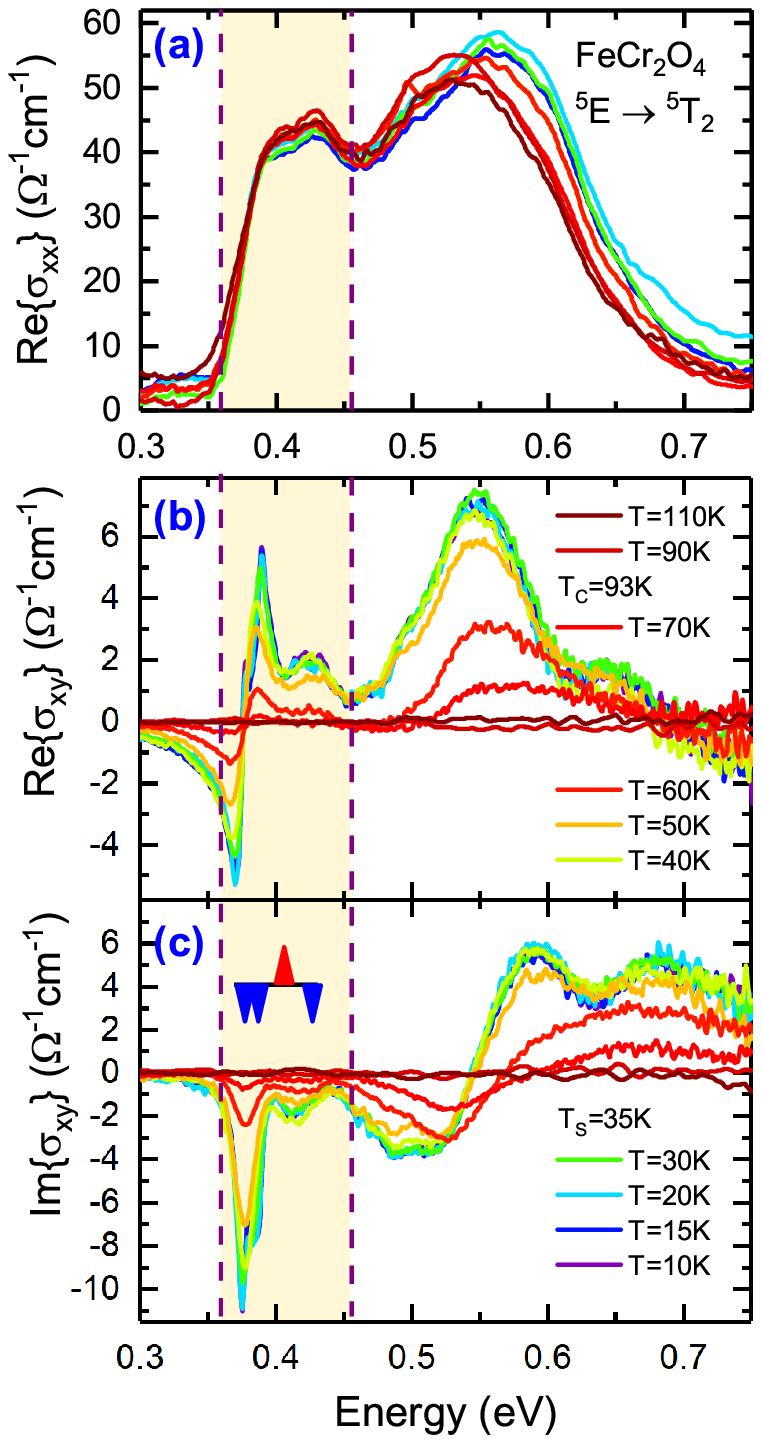}
    \caption{(Color online) Temperature dependence of the diagonal (a) and the complex off-diagonal (b,c) optical conductivity spectra of FeCr$_2$O$_4$. Approximated region of the electronic states of the free-ion spin-orbit split $^5$T$_2$ multiplet is indicated by the coloured stripes. Arrows within the region marks the possible assignment for SOI-split ZPL transitions, the outside region corresponds to the phonon sidebands.}
    \label{moke03}
    \end{figure}
    \end{center}

\subsection{Diagonal and off-diagonal optical conductivity spectra}

Figures~\ref{moke04} and \ref{moke03} show  the off-diagonal optical conductivity spectra along with
the real part of the diagonal optical conductivity for CoCr$_2$O$_4$
and FeCr$_2$O$_4$, respectively. The diagonal
optical conductivity spectrum sums up the
contributions from transitions excited by right and left
($\Delta{J}_z=\pm{1}$) circularly polarized photons as well as non circular contributions ($\Delta{J}_z=0$).
In contrast, the off-diagonal optical
conductivity corresponds to the difference between the contributions
coming from left and right circularly polarized photons, hence it
can have both positive or negative signs.

The temperature dependence of the off-diagonal optical conductivity
has similar evolution for both compounds.
Below $T_{\rm C}$ magnitude of the signal gradually increases
with decreasing temperature and saturates at about 30\,K roughly
following the temperature dependence of the
magnetization.\cite{Ohgushi2008,Yamasaki2006,Shirane1964}
Lifetime of the excited states
also increases towards lower temperatures as clearly manifested in
the decrease of the line widths in both the diagonal and
off-diagonal conductivity spectra.

In CoCr$_2$O$_4$, there are two bands of excitations centered around
0.8\,eV and 2\,eV, which are assigned respectively to the $e^4 t_2^3$
$^4$A$_2$ $\rightarrow$ $e^3 t_2^4$ $^4$T$_2$ and $e^4 t_2^3$ $^4$A$_2$
$\rightarrow$ $e^3 t_2^4$ $^4$T$_1$ on-site electronic transitions\cite{sidenote1} of the
Co$^{2+}$ ions with tetrahedral coordination.
Here we note, that although the low energy $^4$A$_2\rightarrow^4$T$_2$ transition is electric-dipole forbidden
(as A$_2\otimes$T$_2$=\,T$_1$ does not contain the T$_2$ irreducible representation, which is connected to the electric dipole allowed transitions), the spin-orbit coupling eventually turns it electric-dipole allowed.
This is owing to the fact that the SOI splits both the $^4$T$_1$ and $^4$T$_2$ multiplets to the same electronic states.
As a further demonstration, the spectral weight in the diagonal spectra ($\propto\int_{\omega_{min}}^{\omega_{max}}\sigma_{xx}(\omega)d\omega$), of the $^4$A$_2\rightarrow^4$T$_2$ transition allowed by SOI is smaller by a factor of 5 than that of the $^4$A$_2\rightarrow^4$T$_1$ band.
Besides their different magnitudes, there is a sign difference between the
off-diagonal conductivity of the two $d$-$d$ bands as it has been
predicted for the irreducible representations of the different excited ($^4$T$_1$ and $^4$T$_2$)
states in the framework of the crystal field theory.\cite{Sugano1970Book}

In FeCr$_2$O$_4$, similarly to FeCr$_2$S$_4$, there is a single branch of excitation is centered
at about 0.4\,eV \cite{Ohgushi2005}. This excitation has been assigned to the $e^3 t_2^3$
$^5$E $\rightarrow$ $e^2 t_2^4$ $^5$T$_2$ transitions of the tetrahedrally coordinated Fe$^{2+}$ ions.

\subsection{Temperature dependence of the magneto-optical effect}

\begin{center}
    \begin{figure}

    \includegraphics[width=8truecm]{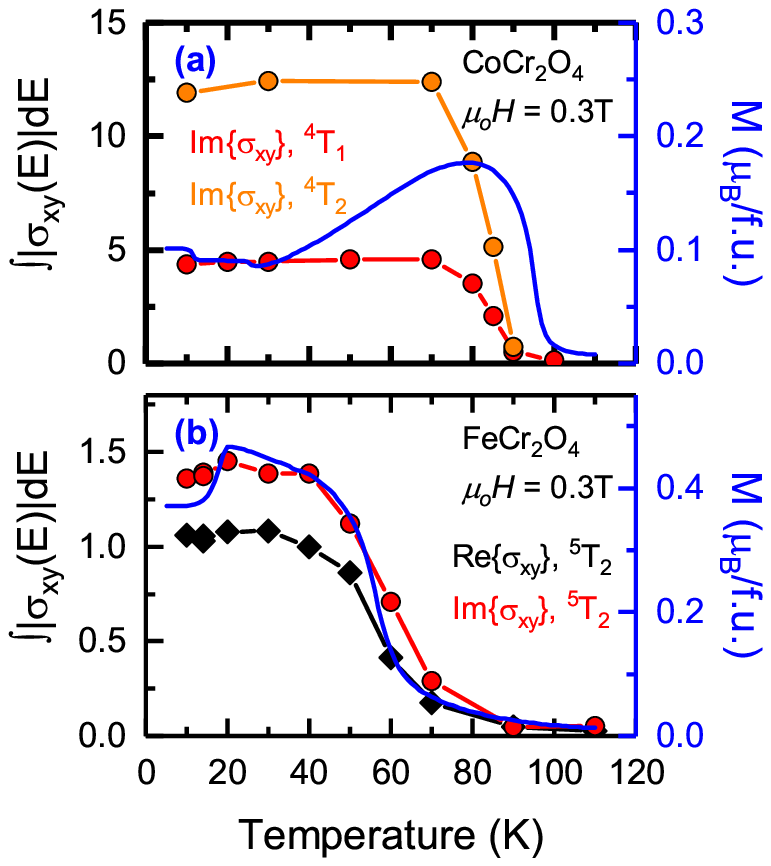}
    \caption{(Color online) Temperature dependence of the magneto-optical effect compared to the magnetization of CoCr$_2$O$_4$ (a) and FeCr$_2$O$_4$ (b). Strength of the magneto-optical effect is represented by the integrals of the absolute values of the real and imaginary parts of the off-diagonal conductivity over the spectral window of the measurement. Magnetization was measured in $\mu_oH$=0.3\,T magnetic field along the [001] axis of the crystal.}
    \label{moke05}
    \end{figure}
    \end{center}

Figure~\ref{moke05} shows the strength of the magneto-optical effect in
FeCr$_2$O$_4$ and CoCr$_2$O$_4$ as a function of temperature,
represented by integrals of the real and imaginary parts of the
optical conductivity spectra over the spectral window of our study.
Since the off-diagonal conductivity changes sign throughout the
spectral range, the absolute value of its real and imaginary parts
were integrated ($\int\vert\sigma_{xy}(E)\vert{dE}$) as an overall measure of the magneto-optical effect
in the two compounds.
For comparison, the temperature dependence of
the magnetization is also plotted in the two cases. The
magneto-optical effect emerges below the magnetic ordering
temperature and it roughly grows as the magnetization, though considerable deviations between them are
discerned well below $T_{\rm C}$. This may come from the fact that the
magneto-optical signal in our case specifically probes one component of the local
magnetization of the Co or Fe ions, while the overall magnetization
has contributions from the Cr ions as well. The
non-monotonous behavior of the magnetization implies different
temperature dependences of the $A$ and Cr sublattice magnetizations.
On the other hand the magneto-optical signal, which measures the sublattice magnetization of the $A$ ions, monotonously increases below $T_{\rm C}$.

\subsection{Fine structure of the $d$-$d$ transitions}

We turn to the discussion of the fine structure describing both the
diagonal and off-diagonal optical conductivity spectra in comparison
with the results reported for on compounds substitutionally doped
with Co$^{2+}$ or Fe$^{2+}$ in tetrahedral or octahedral sites.\cite{Henry1965,Slack1966,Wittekoek1973,Koidl1973,Pappalardo1961,Hjortsberg1977,Hjortsberg1988,Liehr1958}
In general, there are sharp, distinct features emerging at the low-energy sides of
each bands, with significant temperature dependence.
This sharp and well separated feature is the most prominent for the low-energy side of the T$_2$ multiplet of CoCr$_2$O$_4$ as well discerned in the temperature dependence of both $\sigma_{xx}$ and $\sigma_{xy}$ in Fig.~\ref{moke04}.
At the higher-energy side of
these multiplets there are additional transitions with much larger line
width and less significant temperature dependence. Usually, the parallel analysis of the diagonal and
off-diagonal spectra supports an unambiguous assignment of optical
excitations.
While the diagonal spectra contains the sum of all transitions, namely $\Delta{J}_z=0$ and $\Delta{J}_z\pm{1}$ transitions the off-diagonal contains only the $\Delta{J}_z\pm{1}$ transitions with different signs.
However, the fine structure of the $d$-$d$ bands due to strong overlap between these broad
transitions do not allow precise assignment of the excitations.

Width of the $d$-$d$ bands are
$\delta{E}\approx$\,0.3\,eV and $\delta{E}\approx$\,0.5\,eV for
the $^4$A$_2\rightarrow^4$T$_2$ and $^4$A$_2\rightarrow^4$T$_1$
transitions of the Co$^{2+}$ ion, respectively, while $\delta
E\approx$\,0.4\,eV for the $^5$E$\rightarrow^5$T$_2$ transitions of the Fe$^{2+}$ ion.
On the other hand, bandwidths reported for the $d$-$d$ transitions of Co$^{2+}$ and Fe$^{2+}$
ions~\cite{Pappalardo1961,Hjortsberg1977,Hjortsberg1988,Wittekoek1973,
Slack1966,Ohgushi2008} in general are in the range of $\delta{E}\approx$\,0.1-0.3\,eV.
The observed large bandwidths cannot be explained exclusively by
the effect of first-order spin-orbit coupling, since the free ion
spin-orbit interaction parameters are only
$\zeta_o$=410\,cm$^{-1} \approx$\,50\,meV for the $^5$T$_2$
multiplet of Fe$^{2+}$ and $\zeta_o$=533\,cm$^{-1} \approx$70\,meV
for both the $^4$T$_1$ and $^4$T$_2$ multiplets of Co$^{2+}$.
Moreover, strength of the SOI should be further reduced in crystals due to hybridization with the
ligand orbitals.\cite{Ohgushi2008,Sugano1970Book,Jorgensen2013Book,Burns1970Book} The
hypothetic broadening of the $d$-$d$ bands, which would correspond
to the free-ion spin-orbit interaction parameters, are indicated in
Figs.~\ref{moke04} and \ref{moke03} by the highlighted yellow
regions.

\begin{center}
    \begin{figure*}

    \includegraphics[width=16truecm]{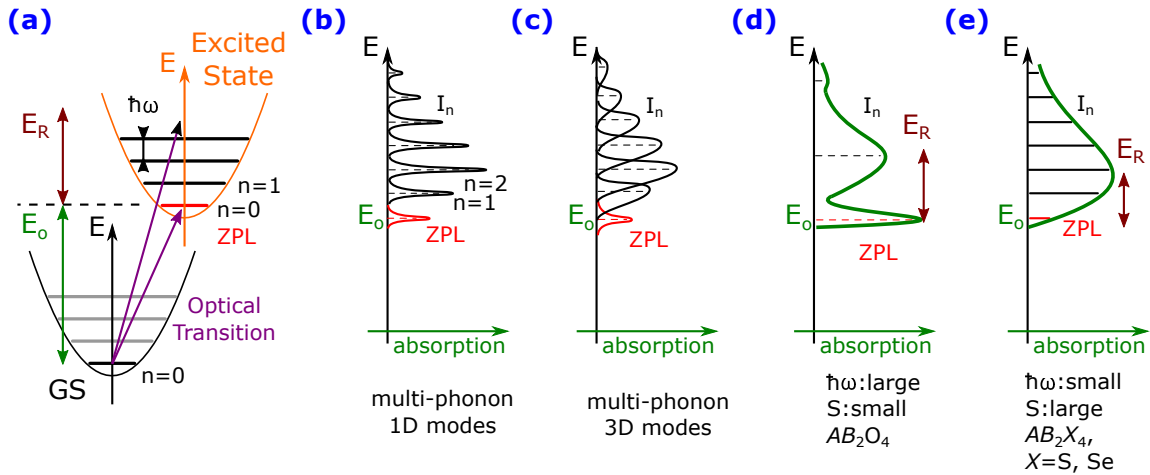}
    \caption{(Color online) (a) Schematic illustration of the hybrid electron-lattice excitations in the presence of electron-phonon coupling. Transitions are allowed from the ground state to the phonon side bands of the excited multiplet, which increases the spectral width of the transition. (b-e) Illustration of the absorption spectra for different models: (b) Dilute systems have spiky spectrum  while (c) regular crystals have wider phonon side bands. In oxide spinels the $\hbar\omega$ phonon energy is larger and the $S$ coupling parameter (Eq.~(\ref{eq_02})) is smaller as compared to chalcogenide spinels. Therefore, in the former case, shown in panel (d), the ZPL is less overlapped by the phonon sidebands while in the second case, sketched in panel (e), the phonon sidebands are not just more overlapped with the ZPL, but also most of the intensity is transferred to the sidebands.}
    \label{moke06}
    \end{figure*}
    \end{center}

In dilute systems with few tetrahedrally coordinated $A^{2+}$ ions,
the increased bandwidth is explained by two main factors: i) the
second order spin-orbit interaction, which slightly further splits the
orbital degeneracy of the $d$ states\cite{Slack1966} and ii) the
emergence of phonon sidebands at the high-energy side of the zero
phonon $d$-$d$ multiplets due to the dynamic Jahn-Teller
effect.\cite{Slack1966,Wittekoek1973,Koidl1973,Pappalardo1961,Hjortsberg1977,Hjortsberg1988,Liehr1958,Toyozawa1966}
The dynamic Jahn-Teller effect is an interplay between the purely
orbital excitations of an ion (zero-phonon line, ZPL) and the
vibrational modes of the surrounding lattice. Due to the
hybridization of these excitations, part of the spectral weight is
transferred from the ZPL to the phonon sidebands (see Fig.~\ref{moke06}).
In most dilute systems, multiplets of the $A^{2+}$ ion are mostly
coupled to phonon modes with one dimensional representation 
(typically to the A$_1$ breathing mode of the ligand cage).
The normalized intensity of the $n$-th sideband is well described by a Poisson distribution~\cite{Toyozawa2003Book}:
\begin{equation}
I_n = e^{-S}\frac{S^n}{n!},\label{eq_02}
\end{equation}
where $S = \frac{E_{\rm R}}{\hbar\omega}$ is the strength of the
electron-phonon coupling, $E_{\rm R}$ is the energy shift of the highest intensity side band, $\hbar\omega$ is the energy of the coupled phonon mode and $I_0 = e^{-S}$ is the intensity of the ZPL.
The expectation number of the phonons coupled to the electronic excitation equals to the strength of the coupling ($S=\langle{n}\rangle$).
Therefore, in dilute systems, the local vibrations of the ligands surrounding the $A^{2+}$ ions
give sharp phonon sidebands to the the optical spectra, which is schematically illustrated in Fig.~\ref{moke06}(b).

In high-symmetry crystals with a regular network of $A^{2+}$ ions where several different phonons with higher dimensional representation may couple to the electronic excitations.
In this case Eq.~(\ref{eq_02}) is expanded to multi-dimensional form and the phonon sidebands are widened\cite{Toyozawa2003Book,Toyozawa1955,Huang1951,Lax1952}, which is illustrated in Fig.~\ref{moke06}(c).
Here we note that the intensity of the ZPL highly depends on the $S$ coupling constant.
For $S=1$ the ZPL is the strongest and the phonon sidebands are evanescent, while for large coupling the ZPL is negligible and the phonon sidebands dominate the spectrum.

Strength of the electron-phonon coupling in spinel compounds can be estimated by determining $E_{\rm R}$ and $\hbar\omega$ separately.
$E_{\rm R}\approx$\,0.15\,eV is well approximated by the width of the multiplets ($2E_{\rm R}=\delta{E}\approx$\,0.3\,eV), which is in general appropriate for both oxide and chalcogenide spinels.\cite{Ohgushi2008}
Phonon energy of the coupled modes can be estimated by the highest energy optical phonon mode, as it is mainly composed of lattice vibrations of the lighter ligand nuclei.
Therefore, for oxide spinels the highest energy optical phonon modes ($\hbar\omega\approx$\,74\,meV) brings a relatively small coupling constant ($S$=2).\cite{Kocsis2013} 
On the other hand, in sulfur and selenium based spinels the ligands are heavier ($\hbar\omega\approx$\,49\,meV and $\hbar\omega\approx$\,37\,meV, respectively\cite{Rudolf2007}) and the electron-phonon couplings are also higher, $S$=3 and $S$=4, respectively.
In sulfide and selenide spinels the expansive $3p$ and $4p$ orbitals of the ligand ions have stronger hybridization with the $3d$ orbitals of the $A^{2+}$ transition metal ions, thus the stronger electron-phonon coupling is expected, in accord with the observations.

\subsection{Criteria of strong magneto-optical effects}

Based on our experimental results, we list the criteria of strong magneto-optical effects at on-site $d$-$d$
transitions of transition metal ions in magnetic crystals.
Besides the well-known criteria (i)-(v), we add two empirical, and material specific ones:

\begin{enumerate}[(i)]
\item To make the on-site $d$-$d$ transitions electric-dipole allowed,
the inversion symmetry at the transition metal sites needs to be
broken, e.g. by the tetrahedral coordination of the sites.

\item Since spin forbidden transitions are usually weak, the on-site $d$-$d$ transitions have to be spin allowed
as well. This excludes the ions with half-filled $d$ shell carrying
spin $S$=5/2.

\item The material has to be a semiconductor with sufficiently large
gap so that the $d$-$d$ transitions lie within the optical band gap.

\item Strong spin-orbit interaction is needed to split the degeneracy
of the $d$ orbitals within the crystal-field multiplets ($e$ and $t_2$ states in spinels with cubic symmetry).

\item The material has to be either a ferro- or ferrimagnet, where the
magnetic exchange interaction removes the degeneracy of the states
with different $J_z$ values.

\item Among the SOI-split multiplets of the excited states, there should be only one with higher spin degeneracy than the ground state. In this case, sum of the oscillator strengths of the remaining transitions to the $\vert{J_z}\vert$-1 excited states will have the same contribution to the magneto-circular birefringence as the single transition to the $\vert{J_z}\vert$+1 state.

\item The ligand ions should be light, so the $\hbar\omega$ is high and the phonon sidebands are well separated from the ZPL. Furthermore, hybridization between the orbitals of the transition metal and the ligand ions should be small. The electron-phonon coupling should be weak, therefore only a small part of the oscillator strength is transferred to the phonon sidebands.

\end{enumerate}

By a systematic analysis of the Tanabe-Sugano
diagrams\cite{Sugano1970Book} and taking into account the effect of
spin-orbit coupling, we found a few transition-metal ions with
tetrahedral environment which might have compliance with these
conditions.
Besides the interesting $e^4 t_2^3$ $^4$A$_2$ ground and the $e^3 t_2^4$ $^4$T$_2$ excited states combination of Co$^{2+}$, the V$^{3+}$ ion has similarly favorable electron configuration.
Experimental studies on vanadium doped compounds have shown that tetrahedrally coordinated V$^{3+}$ ions usually realize $e^2$ $^3$A$_2$ ground and the $e^1 t_2^1$ $^3$T$_1$ excited states, which could be optimal for strong magneto-optical effects.\cite{Koidl1973,Kuck1999}
However, V$^{3+}$ ions, similarly to the Cr$^{3+}$ ions, dominantly occupy octahedral
environments when forming regular sublattice in crystals and only
enter tetrahedral sites as
dopants.\cite{Kuck1999,Sviridov1981} Thus, they are unfavored by
condition (i). Tetrahedrally coordinated Mn$^{+2}$ ions carry $S$=5/2 spins, hence, a strong magneto-optical response from these ions  is excluded by condition (ii). Optical
excitations of Fe$^{2+}$ ions in tetrahedral coordination do not
fulfill condition (vi) as observed in former
works\cite{Kocsis2013,Ohgushi2008,Ohgushi2005,Slack1966,Wittekoek1973,Hjortsberg1977,Hjortsberg1988} as well as in the present
study.
Tetrahedrally coordinated Ni$^{2+}$ and Cu$^{2+}$ ions show orbital degeneracy, which is lifted by the distortion of oxygen tetrahedra upon the cooperative Jahn-Teller transition.\cite{Siratori1967,Crottaz1997,Kennedy2008} In the distorted state, for both ions there is no orbital degeneracy either in the ground or the excited states. Correspondingly, we expect a weak magneto-optical response for these ions.
Therefore, the best candidates which can
generate strong magneto-optical effects are the tetrahedrally coordinated
Co$^{2+}$ ions.

\section{Conclusions}
We observed strong magneto-optical Kerr effect at the on-site $d$-$d$
transitions of Co$^{2+}$ and Fe$^{2+}$ ions in the spinel oxides
CoCr$_2$O$_4$ and FeCr$_2$O$_4$. The magneto-optical Kerr rotation
$\vartheta_{\rm Kerr}=$12\,deg observed in CoCr$_2$O$_4$ is the largest reported so
far in magnetic semiconductors and insulators. We discussed the
criteria of strong magneto-optical effects in insulating transition-metal oxide compounds and the unique potential of the tetrahedrally coordinated Co$^{2+}$ ions.

\begin{acknowledgments}
The authors are grateful to N. Nagaosa, T. Arima, and K. Penc for fruitful
discussions and \'A. Butykai for the technical support in magnetization measurements. This work was supported by the Hungarian Research Funds
OTKA K 108918, OTKA PD 111756 ANN 122879 and Bolyai 00565/14/11 and by the Deutsche Forschungsgemeinschaft through the Transregional Collaborative Research Center TRR 80.
\end{acknowledgments}


\begin{thebibliography}{99}
\bibitem{sidenote1}
	The $e^nt_2^m$ $^S\Gamma$ notation means that $\Gamma$ irreducible representation of a transition metal multiplet, which has $S$ spin degeneracy and the orbital part is derived from the product of $n$ $e$ and $m$ $t_2$ crystal field split orbitals. In the text the electron configuration will be omitted for simplicity.
\end{thebibliography}
\end{document}